\documentclass{iau}
\usepackage{graphicx}

\title[] %% give here short title %%
{Thermophysical Characteristics of OSIRIS-REx Target Asteroid (101955) Bennu}

\author[]   %% give here short author list %%
{Liangliang Yu$^{1,2}$
 \and Jianghui Ji$^1$}

\affiliation{
$^1$Key Laboratory of Planetary Sciences, Purple Mountain Observatory, Chinese Academy of Sciences, Nanjing 210008, China; \\
$^2$Lunar and Planetary Science Laboratory, Macau University of Science and Technology, Taipa, Macau \\
email: {\tt jijh@pmo.ac.cn, yullmoon@pmo.ac.cn} }

\pubyear{2015}
\volume{318}  %% insert here IAU Symposium No.
\jname{Asteroids: New Observations, New Models}
\editors{Steven Chesley, Alessandro Morbidelli \& Robert Jedicke, eds.}

\begin{document}

\maketitle

\begin{abstract}

In this work, we investigate the thermophysical properties, including
thermal inertia, roughness fraction and surface grain size of OSIRIS-REx
target asteroid (101955) Bennu by using a thermophysical model with the
recently updated 3D radar-derived shape model
(\cite[Nolan et al., 2013]{Nolan2013}) and mid-infrared observations
(\cite[M$\ddot{u}$ller et al, 2012]{Muller2012}, \cite[Emery et al., 2014]{Emery2014}).
We find that the asteroid bears an effective diameter of $510^{+6}_{-40}$ m,
a geometric albedo of $0.047^{+0.0083}_{-0.0011}$, a roughness fraction of
$0.04^{+0.26}_{-0.04}$, and thermal inertia of $240^{+440}_{-60}\rm~Jm^{-2}s^{-0.5}K^{-1}$
for our best-fit solution. The best-estimate thermal inertia suggests that fine-grained
regolith may cover a large portion of Bennu's surface, where a grain size may vary from
$1.3$ to $31$~mm. Our outcome suggests that Bennu is suitable
for the OSIRIS-REx mission to return samples to Earth.

%% add here a maximum of 10 keywords, to be taken form the file <Keywords.txt>
\end{abstract}

\firstsection % if your document starts with a section,
              % remove some space above using this command.
\section{Introduction}
(101955) Bennu is an Apollo-type near-Earth asteroid (NEA) given its orbital
characteristics. Bennu is a potential Earth impactor with a relatively high
impact probability of approximately $3.7\times10^{-4}$
(\cite[Milani et al., 2009]{Milani2009}, \cite[Chesley et al., 2014]{chesley2014}).
Recently, \cite{Chesley2014} showed that the semimajor axis of Bennu drifts at an
averaged rate $da/dt=(-19.0\pm0.1)\times10^{-4}~\rm au\cdot Myr^{-1}$ due to the
Yarkovsky effect, and further predicted dozens of potential impacts for this asteroid
from 2175 to 2196. Since its orbit makes it especially accessible for the spacecraft,
Bennu is considered as one of the potentially hazardous asteroids (PHA) and was chosen as
a suitable target for NASA's OSIRIS-REx sample return mission
(\cite[Lauretta et al., 2012]{Lauretta2012}).
The OSIRIS-REx spacecraft is scheduled to be launched in 2016.

Several researchers investigated the thermophysical features of Bennu.
\cite{Muller2012} derived Bennu's thermal inertia of
$\sim$ $650\rm~Jm^{-2}s^{-0.5}K^{-1}$ with thermophysical model,
based on observations from Herschel/PACS, ESO-VISIR, Spitzer-IRS and
Spitzer-PUI. Recently, \cite{Emery2014} showed an update
thermal inertia of Bennu approximately $310\pm70\rm~Jm^{-2}s^{-0.5}K^{-1}$
from a thermophysical analysis of Spitzer-IRS spectra and a multi-band thermal
lightcurve. There are two main different aspects between the work of
(\cite[Emery et al., 2014]{Emery2014}) and
(\cite[M$\ddot{u}$ller et al, 2012]{Muller2012}):
first, the former considered a 3D radar-derived shape model by
(\cite[Nolan et al., 2013]{Nolan2013}) in the modeling process, whereas
the latter adopted a simple spherical shape model; second, the mid-infrared
observations they used were not idenfical, in that IRAC and IRS peak-up data
were included in \cite{Emery2014}, but not utilized in
(\cite[M$\ddot{u}$ller et al, 2012]{Muller2012}).

In this paper, we adopt independently developed thermophysical
simulation codes from (\cite[Yu, Ji \& Wang 2014]{Yu2014}) based on the Advanced Thermal
Physical Model (ATPM) (\cite[Rozitis \& Green 2011]{Rozitis2011}), to investigate the
surface thermophysical characteristics of Bennu. In our modelling process,
we utilize Bennu's radar-derived shape model given by
(\cite[Nolan et al., 2013]{Nolan2013}) rather than a spherical approximation shape
(\cite[M$\ddot{u}$ller et al, 2012]{Muller2012}). Moreover,
we added the mid-infrared data gathered from four groups of observations, at various
phase angles, by Spitzer-PUI, Spitzer-IRAC, Herschel/PACS and ESO VLT/VISIR
(\cite[M$\ddot{u}$ller et al, 2012]{Muller2012}, \cite[Emery et al., 2014]{Emery2014}).
By fitting all observations, we obtain a thermal inertia of Bennu that is slightly
lower than that of (\cite[Emery et al., 2014]{Emery2014}), which indicates
an important evidence of the fine-grained regolith on Bennu's surface
(for details see \cite[Yu \& Ji 2015]{Yu2015}). Moreover, using the derived thermal
inertia, we estimate the grain size of the regolith from a thermal
conductivity model of (\cite[Gundlach \& Blum 2013]{Gundlach2013}).

\section{Analysis}

\begin{figure}
\includegraphics[scale=0.8]{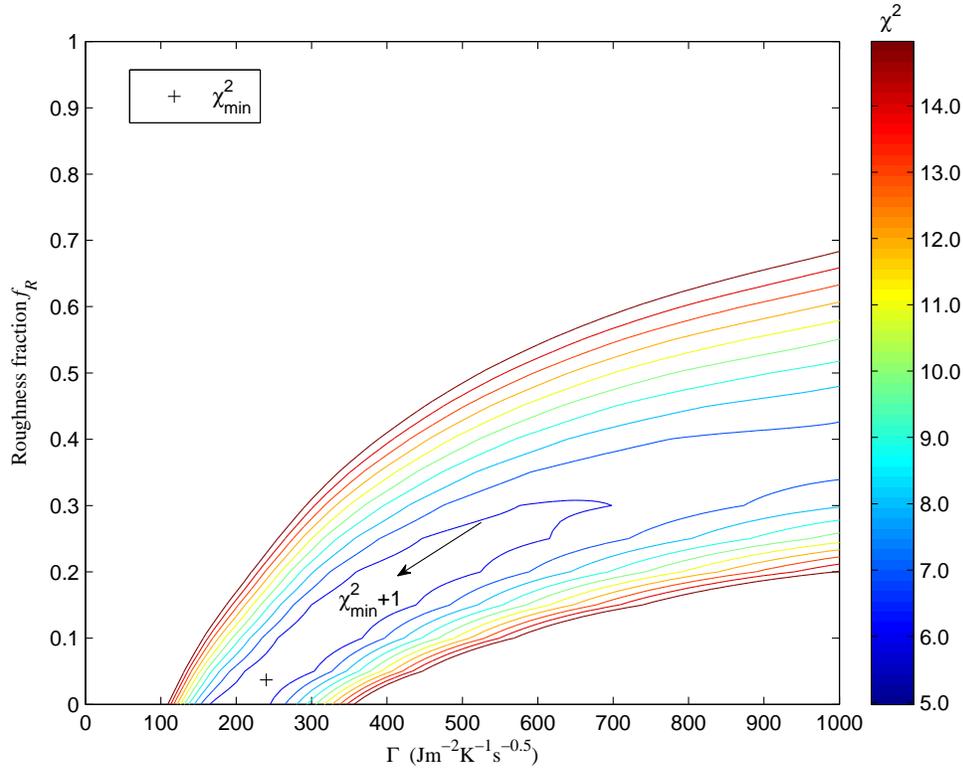}
  \centering
  \caption{$\chi^{2}$ ($\Gamma$, $f_{\rm R}$) contour according to ATPM results.
  The color (from blue to red) relates to the increase of profile of $\chi^{2}$.
  The blue curve labeled by $\chi^{2}_{\rm min}+1$ is taken as the 1$\sigma$
  limit to the free fit parameters
  (\cite[Emery et al., 2014]{Emery2014},\cite[Bevington \& Robingson 2003]{Bevington2003}
  ).
  }\label{chi2contour}
\end{figure}

Figure \ref{chi2contour} shows a contour of $\chi^{2}$ in the 2-dimensional
parameter space ($\Gamma$, $f_{\rm R}$), where the increase of $\chi^{2}$ is
shown by color bar from blue to red. The black '+' corresponds to the minimum
$\chi_{\rm min}^{2}$. And in this case, we have
$\Gamma=240\rm~Jm^{-2}s^{-0.5}K^{-1}$ and $f_{\rm R}=0.04$, as best
fit solutions to the observations. The blue curve is the profile of $\chi_{\rm min}^{2}+1$,
which is assumed as the 1$\sigma$ limit of the free fit parameters $\Gamma$ and $f_{\rm R}$.
The blue profile forms a closed region in the ($\Gamma$, $f_{\rm R}$) space. Thus,
we can simultaneously constrain thermal inertia and roughness fraction in 2-dimensional space,
considering the 1$\sigma$ limit. However, the contour curves of $\chi^{2}$ above 1$\sigma$
cannot froms closed regions, suggesting that the degeneracy of thermal inertia and roughness fraction
cannot be removed so well like the 1$\sigma$ level, thus thermal inertia and roughness
fraction may be simply separated at the 1$\sigma$ level based on the ATPM calculations.
Therefore, if the 1$\sigma$ limit is reliable, we may safely conclude that the roughness
fraction is possible in the range of 0$\sim$0.3, and the thermal inertia may be
in the range of $180\sim680\rm~Jm^{-2}s^{-0.5}K^{-1}$. Our results agree
with both earlier investigations of
(\cite[M$\ddot{u}$ller et al, 2012]{Muller2012}) and
(\cite[Emery et al., 2014]{Emery2014}).

With the above derived thermal inertia, the grain size of Bennu can be estimated
according to the method of (\cite[Gundlach \& Blum 2013]{Gundlach2013}).
A mean surface temperature $T=300$~K is assumed in our computation. The other
parameters are taken from (\cite[Gundlach \& Blum 2013]{Gundlach2013}).
Based on the thermal inertia $\Gamma=240\rm~Jm^{-2}s^{-0.5}K^{-1}$, the grain
radius is likely to be in the range $2\sim5$~mm. In addition, the grain radius
may be estimated as ranging from $1.3$ to $\sim31$~mm considering  1$\sigma$ range
of thermal inertia. On the basis of this estimation of grain radius, we infer
that boulders or rocks may be few on the surface of Bennu, suggesting that the
Touch-And-Go Sample Acquisition Mechanism (TAGSAM) designed by the OSIRIS-REx
team should find an accessible environment to operate successfully.

\section*{Acknowledgments}
This work is financially supported by National Natural Science
Foundation of China (Grants No. 11273068, 11473073, 11403105), the Strategic
Priority Research Program-The Emergence of Cosmological Structures
of the Chinese Academy of Sciences (Grant No. XDB09000000), the
innovative and interdisciplinary program by CAS (Grant No.
KJZD-EW-Z001),  the  Natural Science Foundation of  Jiangsu  Province
(Grant  No. BK 20141509), and the Foundation of Minor Planets of
Purple Mountain Observatory.

\end{document}